\documentclass[prd,onecolumn]{revtex4}
\usepackage{dcolumn}
\usepackage{multirow}
\usepackage{graphicx}
\usepackage{amssymb}
\usepackage{bm}
\usepackage{hyperref}
\usepackage{epstopdf}
\usepackage{color}
\usepackage{mathrsfs}
\usepackage{amsmath,amssymb,amsthm}
\begin{document}
\title{Thin-shell wormholes constrained by cosmological observations}
\author{Deng Wang}
\email{Cstar@mail.nankai.edu.cn}
\affiliation{Theoretical Physics Division, Chern Institute of Mathematics, Nankai University,
Tianjin 300071, China}
\author{Xin-He Meng}
\email{xhm@nankai.edu.cn}
\affiliation{{Department of Physics, Nankai University, Tianjin 300071, China}\\
{State Key Lab of Theoretical Physics,
Institute of Theoretical Physics, CAS, Beijing 100080, China}}
\begin{abstract}
We investigate the thin-shell wormholes constrained by cosmological observations for the first time in the literature. Without loss of generality, we study the thin-shell wormholes in $\omega$CDM model and analyze their stability under perturbations preserving the symmetry. Firstly, we constrain the $\omega$CDM model using a combination of Union 2.1 SNe Ia data, the latest $H(z)$ data and CMB data. Secondly, we use the constrained dark energy equation of state (EoS) $\omega$ which lies in $[-1.05,-0.89]$ to investigate thin-shell wormholes of various black hole spacetimes. We find that the stable Schwarzschild and Reinssner-Nordstr\"{o}m thin-shell wormholes constrained by cosmological observations do not exist. In addition, the method we developed can be applied into exploring the stable thin-shell wormholes from any black hole spacetime in the framework of any cosmological theory.
\end{abstract}
\maketitle
\section{Introduction}
In recent years, the seminal paper by Morris and Thorne \cite{1} has motivated a great deal of studies to explore the traversable Lorentzian wormholes. As exotic solutions of gravitational theories, these objects have a throat that connects two widely separated regions of the same universe or two different universes \cite{1,2}. In general theory of relativity (GR), traversable wormholes are supported by exotic matter that violates the null energy condition (NEC). Using the '' volume integral quantifier '' (VIQ) \cite{3}, the amounts of exotic matter can be substantially small, but at the expense of large stresses at the throat. Further discussions about the energy conditions of wormhole spacetime configurations can be found in Ref. \cite{4}.

Generally speaking, thin-shell wormholes are constructed by the so-called cut-and-paste technique, i.e., grafting together two black hole spacetimes at the junction surface which corresponds to a three-dimensional thin shell \cite{2,5,6}. These thin-shell wormholes have been widely investigated in the literature, since the stability analysis is very easy to implemented and the exotic matter is only confined to the thin shell. The stability analysis of spherically symmetric thin-shell wormholes have been implemented under the radial perturbations that preserves the symmetries by several authors \cite{7,8,9,10,11,12}. The thin-shell wormholes with cylindrical, plane and conformal symmetries have already been analyzed in Refs. \cite{13,14}. The Schwarzschild thin-shell wormholes with variable equation of state (EoS) was also taken into consideration in Ref. \cite{15}.

In the past few years, a renewed interest has arisen in the old field---wormholes, since the elegant discovery that our universe is undergoing a phase of an accelerated expansion (in both cases, the NEC is violated) \cite{16,17}. Dark energy as new source of exotic matter has inspired a number of studies to explore the corresponding wormhole geometries in various kinds of cosmological theories. For instance, wormholes supported by phantom energy, Chaplygin gas, generalized Chaplygin gas, viscous fluid and Shan-Chen fluid have been investigated vividly in Refs. \cite{18,19,20,21,22,23}. Recently, with more and more high-quality cosmic data, we have studied the geometrical and holographical dark energy wormholes constrained by astrophysical observations for details, and verified that the exotic spacetime configurations wormholes can actually exist in the universe \cite{24,25}. Based on this concern, a question naturally comes into being, namely, whether one can similarly investigate the thin-shell wormholes constrained by cosmological observations ? Note that this is the starting point of the present work.

This paper is organized in the following manners. In Section 2, we place constraints on the simple $\omega$CDM model using a combination of Union 2.1 type Ia supernovae (SNe Ia) data, the latest $H(z)$ data and the cosmological microwave background (CMB) measurements. In Section 3, we make a review on the thin-shell wormhole formalism. In Section 4, we use the constrained cosmological parameter to study the thin-shell wormholes constructed by various spacetimes. The discussions and conclusions are presented in the final section (we use units $G=c=1$ throughout the context).

\section{statistical constraints}
For the purpose to constrain the $\omega$CDM model better, we express its dimensionless Hubble parameter $E(z)$ as follows
\begin{equation}
E(z)=[\Omega_{m0}(1+z)^3+(1-\Omega_{m0})(1+z)^{3(1+\omega)}]^{\frac{1}{2}},   \label{1}
\end{equation}
where $z$ denotes the redshift, $\Omega_{m0}$ the matter density ratio parameter and $\omega$ the dark energy EoS. Subsequently, we would like to exhibit our constraining methodology, available cosmic data and constraining results.

\subsection{Type Ia Supernovae}
The observations of SNe Ia provide an powerful tool to explore the expansion history of the universe. It is well known that the absolute magnitudes of all the SNe Ia are considered to be the same, since all SNe Ia almost explode at the same mass ($M\approx-19.3\pm0.3$). Based on this concerning, in theory SNe Ia can be used as the standard candles. In this study, we take the Union 2.1 data sets without systematic errors for data fitting, consisting of 580 points covering the range of the redshift $z\in(0.015, 1.4)$ \cite{26}. To perform the so-called $\chi^2$ statistics, the theoretical distance modulus is defined as
\begin{equation}
\mu_{th}(z_i)=5\log_{10}d_L(z_i)+\mu_0,,   \label{2}
\end{equation}
where $\mu_0=42.39-5\log_{10}h$, $h$ is the dimensionless Hubble parameter at present in units of 100 $km^{-1}s^{-1}Mpc$. The luminosity distance $d_L(z_i)$ at a given redshift $z_i$ in units of Mpc can be defined as
\begin{equation}
d_L(z_i)=(1+z_i)\int^{z_i}_0\frac{dz'}{E(z';\delta)},   \label{3}
\end{equation}
where $\delta$ represents a set of model parameters. After marginalizing the nuisance parameter $\mu_0$ analytically, the $\chi^2$ for SNe Ia observations can be expressed as
\begin{equation}
\chi^2_{SN}=A(\delta)-\frac{[B(\delta)]^2}{C},   \label{4}
\end{equation}
with
\begin{equation}
A(\delta)=\sum^{580}_{i,j=1}[\mu_{th,i}(z_i;\delta)-\mu_{obs,i}(z_i)]Cov^{-1}_{ij}[\mu_{th,j}(z_j;\delta)-\mu_{obs,i}(z_j)],   \label{5}
\end{equation}
\begin{equation}
B(\delta)=\sum^{580}_{i,j=1}Cov^{-1}_{ij}[\mu_{th,j}(z_j;\delta)-\mu_{obs,i}(z_j)],   \label{6}
\end{equation}
\begin{equation}
C=\sum^{580}_{i,j=1}Cov^{-1}_{ij},   \label{7}
\end{equation}
where $\mu_{obs}$ and $Cov^{-1}_{ij}$ denote the observed distance modulus and the inverse of the covariance metric, respectively.

\subsection{H(z) Measurement}
The recent Hubble parameter measurements have been proven to be very powerful in constraining the cosmological parameters. In the literature, there are two main methods of independent observational $H(z)$ measurement, i.e., `` radial baryon acoustic oscillations (BAO) method '' and `` differential age method ''. In this situation, we use the latest 40 observational Hubble parameter data points, which contains the latest five measurements by Moresco et al. \cite{27} and the catalog in Refs. \cite{27,28}. The corresponding $\chi^2$ for $H(z)$ data is
\begin{equation}
\chi^2_{H}=-\frac{B^2}{A}+C,  \label{8}
\end{equation}
with
\begin{equation}
A=\sum^{40}_{i=1}\frac{E^2(z_i)}{\sigma_i^2}, \label{9}
\end{equation}
\begin{equation}
B=\sum^{40}_{i=1}\frac{E(z_i)H_{obs}(z_i)}{\sigma_i^2}, \label{10}
\end{equation}
\begin{equation}
C=\sum^{40}_{i=1}\frac{H^2_{obs}(z_i)}{\sigma_i^2}, \label{11}
\end{equation}
where $H_{obs}$ and $\sigma_i$ denote the observed Hubble parameter at a given redshift $z_i$ and the corresponding $1\sigma$ statistical error, respectively.

\subsection{CMB Data}
As is well known, the CMB not only provides some useful information on the very early universe, but also can give information on the expansion history of the universe through the positions of the CMB acoustic peaks. More precisely, the likelihood of the shift parameter $\mathcal{R}$ the acoustic scale $l_a$ extracted from the CMB angular power spectrum can allow one to constrain the cosmological parameters \cite{29,30}, where $\mathcal{R}$ determines the overall amplitude of the acoustic peaks and $l_a$ represents the average acoustic peak structure, given by
\begin{equation}
\mathcal{R}=(1+z_\star)H_0\Omega_{m0} D_A(z_\star), \label{12}
\end{equation}
and
\begin{equation}
l_a(z_\star)=(1+z_\star)\frac{\pi D_A(z_\star)}{r_s(z_\star)}, \label{13}
\end{equation}
respectively. Note that here $D_A(z_\star)$ is the comoving angular diameter distance, $r_s(z_\star$ is the comoving sound horizon at the photon-decoupling epoch \cite{31}, and the redshift $z_\star$ for the photon-decoupling is provided by \cite{32}
\begin{equation}
z_\star=1048[1+0.00124(\Omega_{b0}h^2)^{-0.738}][1+g_1(\Omega_{m0}h^2)^{g_2}], \label{14}
\end{equation}
where
\begin{equation}
g_1=\frac{0.0783(\Omega_{b0}h^2)^{-0.238}}{1+39.5(\Omega_{b0}h^2)^{-0.763}}, \qquad g_2=\frac{0.56}{1+21.1(\Omega_{b0}h^2)^{1.81}}. \label{15}
\end{equation}
Here we adopt the CMB data obtained in Ref. \cite{33} through a combination of Planck temperature data, Planck lensing and WMAP polarization: $l_a(z_\star)=301.37$, $\mathcal{R}(z_\star)=1.7407$ and $\Omega_{b0}h^2=0.02228$ with the following inverse covariance matrix
\begin{equation}
C^{-1}_{CMB}=\left(\begin{matrix}
             43.0180 & -366.7718 & 2972.5  \\
             -366.7718 & 24873.0 & 4.4650\times10^5  \\
             2972.5 & 4.4650\times10^5 & 2.1555\times10^7
             \end{matrix}\right). \label{16}
\end{equation}
Then the corresponding $\chi^2$ for CMB data is
\begin{equation}
\chi^2_{CMB}=\Delta p_i(Cov^{-1}_{CMB})\Delta p_j, \qquad \Delta p_i=p^{th}_i-p^{obs}_i, \label{17}
\end{equation}
where $p_1=l_a(z_\star)$, $p_2=\mathcal{R}(z_\star)$ and $p_3=\Omega_{b0}h^2$. Therefore, the $\chi^2$ for the joint constraints by using SNe Ia, H(z) and CMB data sets can be defined as
\begin{equation}
\chi^2_{tot}=\chi^2_{SN}+\chi^2_{H}+\chi^2_{CMB}. \label{18}
\end{equation}

\begin{figure}
\centering
\includegraphics[scale=0.5]{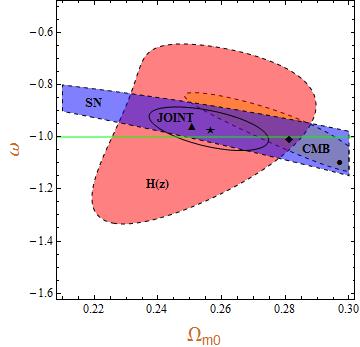}
\caption{The $1\sigma$ confidence intervals for the parameter pair ($\Omega_{m0},\omega$) of $\omega$CDM model constrained by using different data sets. The blue, red, yellow regions and black curve correspond to the $1\sigma$ confidence intervals of SNe Ia, H(z), CMB and joint data sets, respectively. The best fitting points of SNe Ia, H(z), CMB and joint data sets are shown as $\blacklozenge$, $\blacktriangle$, $\bullet$ and $\bigstar$, respectively. The green line represents the base cosmological model.}\label{f1}
\end{figure}

The two-dimensional likelihoods of the parameter pair ($\Omega_{m0},\omega$) of $\omega$CDM model by using different data sets are depicted in Fig. \ref{f1}. One can easily find that the constraining results from SNe Ia, H(z), CMB and joint data sets are all consistent with the the base cosmological model at $1\sigma$ level. Moreover, the best fitting points of the SNe Ia, H(z) and joint data sets are closer to the the base cosmological model than that of CMB data. In the following context, to investigate the thin-shell wormholes supported by astrophysical observations, we would like to adopt the $1\sigma$ range of the dark energy EoS, i.e., $\omega\in[-1.05,0.89]$.

In the previous literature, all the authors analyze the stability of thin-shell wormholes by only choosing the value of $\omega$ arbitrarily. However, in this situation, we give out the possible range of $\omega$ by cosmological observations in order to avoid choosing the value of $\omega$ arbitrarily. Hence, we can exhibit the stability analysis of thin-shell wormholes, which are constructed by different black hole spacetimes.

\section{Thin-shell wormholes}
Take into account a spherically symmetric spacetime geometry
\begin{equation}
ds^2=-f(r)dt^2+f(r)^{-1}dr^2+h(r)(d\theta^2+sin^2\theta d\phi^2), \label{19}
\end{equation}
where $r$ is the radial coordinate which runs in the range $r_0\leq r<\infty$, $0\leq\theta\leq\pi$ and $0\leq\phi\leq2\pi$ are the angular coordinates, and the functions $f(r)$ and $h(r)$ are positive for a given radius (we use $h(r)=r^2$ in the following context). To construct a thin-shell wormhole, we choose a radius $a$ and then remove from each the four-dimensional region:
\begin{equation}
\mathcal{M}^{\pm}=\{r<a|a>r_h\}, \label{20}
\end{equation}
where $r_h$ denotes the maximal event horizon of the black hole. Subsequently, identifying the time-like hypersurface $\partial\mathcal{M}={r=a|a>r_h}$, one can create a new geodesically complete manifold. For a given radius $r$, the area $4\pi r^2$ is minimal when $r=a$, so the manifold corresponds to a wormhole with two regions connected by a throat radius $a$ where the flare-out condition is automatically satisfied. In what follows, using the usual Darmois-Israel formalism [], the Lanczos equations can be expressed as
\begin{equation}
S^i_j=-\frac{1}{8\pi}(\kappa^i_j-\delta^i_j\kappa^k_k), \label{21}
\end{equation}
where $S^i_j$ and $\kappa^i_j$ represent, respectively, the surface stress energy tensor at the junction surface $\partial\mathcal{M}$ and the discontinuity of the extrinsic curvatures across the surface $\partial\mathcal{M}$. Furthermore, in terms of surface energy density $\sigma$ and the surface pressure $\mathcal{P}$, $S^i_j=$diag$(-\sigma,\mathcal{P},\mathcal{P})$. After simple derivations, the energy density $\sigma$ and lateral pressure $\mathcal{P}$ of the thin shell situated at a radius $a$ can be expressed as
\begin{equation}
\sigma=-\frac{\sqrt{f(a)+\dot{a}^2}}{2\pi a}, \label{22}
\end{equation}
and
\begin{equation}
\mathcal{P}=\frac{2a\ddot{a}+2\dot{a}^2+2f(a)+af'(a)}{8\pi a\sqrt{f(a)+\dot{a}^2}}. \label{23}
\end{equation}
It is easy to find that the the weak energy condition (WEC) is violated, since the energy density $\sigma$ on the thin shell  is negative and the radial pressure is zero.

\subsection{Schwarzschild Thin-shell Wormholes}
First of all, we study the thin-shell wormholes constructed by Schwarzschild spacetimes. For this case, the function $f(r)$ reads $f(r)=1-2M/r$, where $M$ is the mass. The energy density $\sigma$ and pressure $\mathcal{P}$ satisfy the conservation equation
\begin{equation}
\frac{d}{d\tau}(\sigma a^2)+\mathcal{P}\frac{d}{d\tau}(a^2)=0, \label{24}
\end{equation}
where $\tau$ is the proper time. In what follows, we would like to use the dark energy EoS $\mathcal{P}=\omega\sigma$ of $\omega$CDM model to investigate the possible static equilibrium configurations, thus, $\dot{a}=\ddot{a}=0$ throughout this work. Furthermore, one can easily obtain the solution of Eq. (\ref{23}) as follows
\begin{equation}
\sigma(a)=\sigma(a_0)\left(\frac{a_0}{a}\right)^{2(1+\omega)}. \label{25}
\end{equation}
Subsequently, by rearranging Eq. (\ref{22}) and defining the potential $V(a)=f(a)-[2\pi a\sigma(a)]^2$, we obtain the equation of motion as
\begin{equation}
\dot{a}^2+V(a)=0. \label{26}
\end{equation}
By expanding $V(a)$ around $a_0$, we have
\begin{equation}
V(a)=V(a_0)+V'(a_0)(a-a_0)+\frac{1}{2}V''(a_0)(a-a_0)^2+O[(a-a_0)^3]. \label{27}
\end{equation}
Since we linearize $V(a)$ around $a=a_0$, the conditions $V(a_0)=0$ and $V'(a_0)=0$ must be satisfied. Furthermore, to obtain a stable equilibrium configuration, the condition $V''(a_0)>0$ is also required. For the Schwarzschild case, after some derivations, we obtain
\begin{equation}
V(a)=1-\frac{2M}{a}-(1-\frac{2M}{a_0})\left(\frac{a_0}{a}\right)^{4\omega+2}. \label{28}
\end{equation}
It is easy to be checked that the first condition $V(a_0)=0$ is satisfied. Then from $V'(a_0)=0$ we get the relation
\begin{equation}
\omega=\frac{1-\frac{a_0}{M}}{\frac{2a_0}{M}-4}. \label{29}
\end{equation}
Since the $1\sigma$ range of dark energy EoS is $\omega\in[-1.05,0.89]$, one can naturally determine the range of stable junction radius
\begin{equation}
2.91<\frac{a_0}{M}<3.28. \label{30}
\end{equation}
Using Eqs. (\ref{29}-\ref{30}), we can have
\begin{equation}
V''(a_0)=-\frac{2M}{a_0^2(a_0-2M)}>0. \label{31}
\end{equation}
Usually speaking, $a_0>2M$ for the Schwarzschild black hole since its event horizon is located at $r_h=2M$. Therefore, the range of junction radius Eq. (\ref{30}) from observational constraints is permitted. However, according to the usual equilibrium relation Eq. (\ref{31}), i.e., $a_0<2M$, we find that the Schwarzschild thin-shell wormholes are unstable from the point views of both theory and observations.

\subsection{Schwarzschild-de Sitter Thin-shell Wormholes}
For the Schwarzschild-de Sitter spacetime, the function has the form $f(r)=1-2M/r-\Lambda r^2/3$, where the cosmological constant $\Lambda>0$. In order to keep $f(r)$ positive, the condition $0<\Lambda M^2\leqslant1/9$ must be satisfied. In this case the geometry has two horizons, i.e., the event and the cosmological ones. Assuming $a$ is greater than the maximal horizon and adopting the same step as the Schwarzschild case, we obtain
\begin{equation}
V(a)=1-\frac{2M}{a}-\frac{\Lambda a^2}{3}-(1-\frac{2M}{a_0}-\frac{\Lambda a_0^2}{3})\left(\frac{a_0}{a}\right)^{4\omega+2}. \label{32}
\end{equation}
It is easy to see $V(a_0)=0$. One can also work out $\omega$ by the condition $V'(a_0)=0$
\begin{equation}
\omega=-\frac{1}{2}\frac{1-\frac{M}{a_0}-\frac{2}{3}\Lambda M^2(\frac{a_0}{M})^2}{1-\frac{2M}{a_0}-\frac{1}{3}\Lambda M^2(\frac{a_0}{M})^2}. \label{33}
\end{equation}
\begin{figure}
\centering
\includegraphics[scale=0.3]{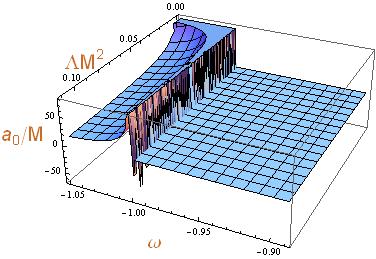}
\includegraphics[scale=0.3]{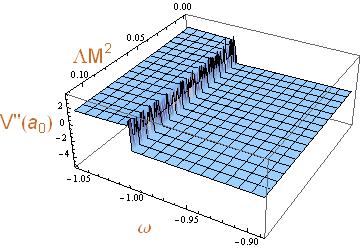}
\includegraphics[scale=0.3]{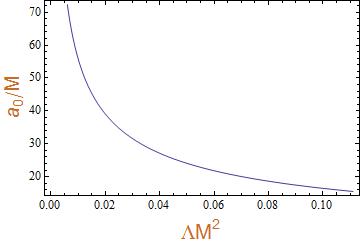}
\includegraphics[scale=0.3]{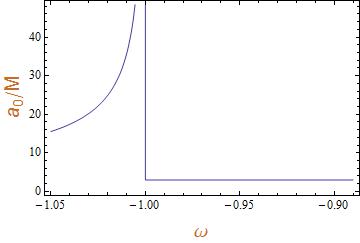}
\caption{Plots for the Schwarzschild spacetime. Left panel: the dependence of $a_0/M$ on the quantities $\omega$ and $\Lambda M^2$. Second panel: the dependence of $V''(a_0)$ on the quantities $\omega$ and $\Lambda M^2$. Third panel: the relation between $a_0/M$ and $\Lambda M^2$. Right panel: the relation between $a_0/M$ and $\omega$.}\label{f2}
\end{figure}

As before, we still use the $1\sigma$ range $\omega\in[-1.05,0.89]$ to determine the possible range of $a_0/M$. Nonetheless, since the its analytic expression is too lengthy, we solve Eq. (\ref{33}) numerically exhibit the behaviors of $a_0/M$ graphically. In the left panel of Fig. \ref{f2}, we show the dependence of $a_0/M$ on the quantities $\omega$ and $\Lambda M^2$, and find that the minimal value of stable junction radius is $a_0/M=15.45$. In the second panel of Fig. \ref{f2}, we exhibit the stability region where $V''(a_0)>0$. One can easily find that the thin-shell wormholes are stable when $\omega<-1$. In the third panel of Fig. \ref{f2}, the relation between $a_0/M$ and $\Lambda M^2$ is shown when $\omega=-1.05$ lies in the $1\sigma$ range $[-1.05,0.89]$. It is easy to see that $a_0/M$ is monotonically decreasing with increasing $\Lambda M^2$. In the right panel of Fig. \ref{f2}, we also exhibit the relation between $a_0/M$ and $\omega$ is shown when $\Lambda M^2=1/9$, and fin that $a_0/M$ diverges at $\omega=-1$ and reaches the minimal value at $\omega=-1.05$.

In summary, for the Schwarzschild-de Sitter case, the stable thin-shell wormholes constrained by cosmological observations must satisfy two conditions: (i) $a_0/M>15.45$; (ii) $\omega\in[-1.05,-1)$.

\subsection{Schwarzschild-anti de Sitter Thin-shell Wormholes}
For the Schwarzschild-anti de Sitter spacetime, the metric function $f(r)$ has the same form as the Schwarzschild-de Sitter case, but with a negative cosmological constant, i.e., $\Lambda >0$. Obviously, the event horizon is situated at
\begin{equation}
r_h=\frac{1-(-3\sqrt{|\Lambda|}M+\sqrt{1+9\Lambda M^2})^{\frac{2}{3}}}{\sqrt{|\Lambda|}(-3\sqrt{|\Lambda|}M+\sqrt{1+9\Lambda M^2})^{\frac{2}{3}})^{\frac{1}{3}}}, \label{34}
\end{equation}
which is a continuous and increasing function of $\Lambda$ and lies in the range $0<r_h<2M$. Differentiating Eq. (\ref{32}) with respect to $a$ and evaluating at $a_0$, we obtain
\begin{equation}
V''(a_0)=\frac{2}{a_0^2}\frac{-\frac{M}{a_0}+3\Lambda M^2(\frac{a_0}{M})-\frac{2}{3}\Lambda M^2(\frac{a_0}{M})^2}{1-\frac{2M}{a_0}-\frac{1}{3}\Lambda M^2(\frac{a_0}{M})^2}>0. \label{35}
\end{equation}
Assuming a negative denominator and using the condition $a_0>2M>r_h$, we find that $\Lambda M^2 >0$, which is not compatible with $\Lambda<0$. The we consider a positive denominator for Eq. (\ref{35}) and derive finally
\begin{equation}
\Lambda M^2<\frac{1}{3(\frac{a_0}{M})^2-\frac{2}{3}(\frac{a_0}{M})^3}.  \label{36}
\end{equation}

\begin{figure}
\centering
\includegraphics[scale=0.3]{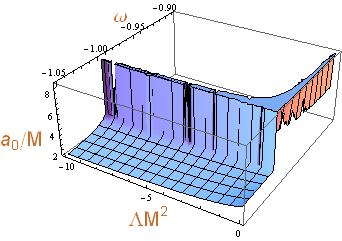}
\includegraphics[scale=0.3]{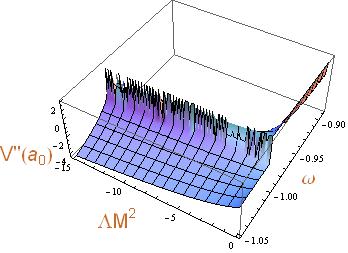}
\includegraphics[scale=0.3]{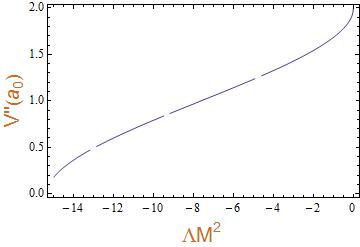}
\includegraphics[scale=0.3]{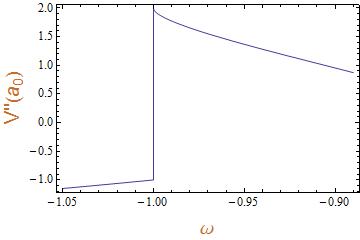}
\caption{Plots for the Schwarzschild-de Sitter spacetime. Left panel: the dependence of $a_0/M$ on the quantities $\omega$ and $\Lambda M^2$. Second panel: the dependence of $V''(a_0)$ on the quantities $\omega$ and $\Lambda M^2$. Third panel: the relation between $a_0/M$ and $\Lambda M^2$. Right panel: the relation between $a_0/M$ and $\omega$.}\label{f3}
\end{figure}
\begin{figure}
\centering
\includegraphics[scale=0.3]{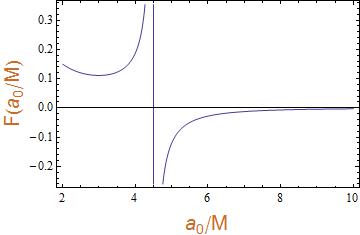}
\caption{The relation between $F(a_0)/M$ and $a_0/M$.}\label{f4}
\end{figure}
To exhibit the stability analysis, we still start from Eq. (\ref{33}) but with a negative $\Lambda$ to show the behaviors of the quantities $a_0/M$ and $V''(a_0)$ graphically. In the left and second panels of Fig. \ref{f3}, we find that, to satisfy $a_0/M>2$ and $ V''(a_0)>0$, the dark energy EoS $\omega$ is restricted in $(-1,-0.89]$ and $\Lambda M^2<0$. In the third panel of Fig. \ref{f3}, the relation between $ V''(a_0)$ and $\Lambda M^2$ when $\omega=-0.998$ is shown, and $V''(a_0)$ is monotonically increasing with increasing $\Lambda M^2$. Moreover, the serrated border of $a_0/M$ and $ V''(a_0)$ in the left and second panels of Fig. \ref{f3} can be well explained by the disconnected parts in the third panel of Fig. \ref{f3}. In the right panel of Fig. \ref{f3}, we also show the relation between $ V''(a_0)$ and $\omega$ when $\Lambda M^2=-0.06$. It is easy to see that $ V''(a_0)<0$ when $\omega<-1$ and $ V''(a_0)>0$ in the range $\omega\in(-1,-0.89]$. Furthermore, to analyze the stability more vividly, we define a new function $F(a_0/M)$ as
\begin{equation}
F(a_0/M)=\frac{1}{3(\frac{a_0}{M})^2-\frac{2}{3}(\frac{a_0}{M})^3}.  \label{37}
\end{equation}
We also exhibit the relation between $F(a_0/M)$ and $a_0/M$ in Fig. \ref{f4}, and find that $F(a_0/M)>0$ when $2<a_0/M<4.5$, $F(a_0/M)<0$ when $a_0/M>4.5$ and $F(a_0/M)$ diverges when $a_0/M=4.5$.
\begin{figure}
\centering
\includegraphics[scale=0.3]{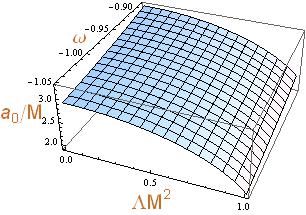}
\includegraphics[scale=0.3]{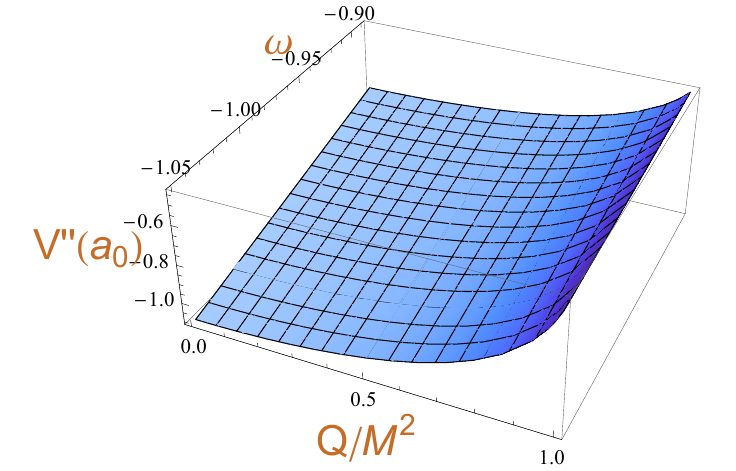}
\includegraphics[scale=0.3]{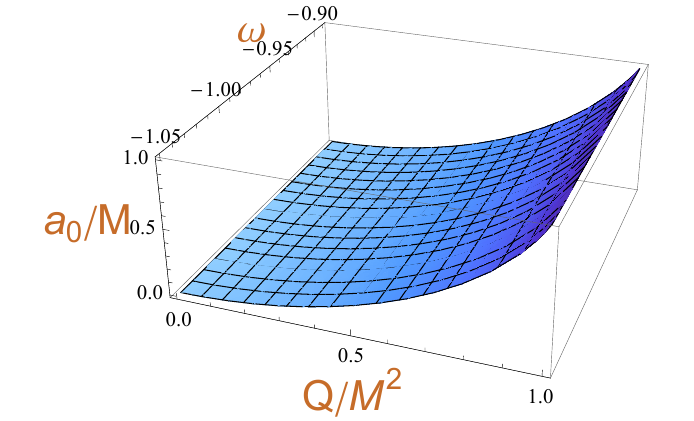}
\caption{Plots for the Reinssner-Nordstr\"{o}m spacetime. Left panel: the dependence of $a_0/M$ on the quantities $\omega$ and $Q^2/M^2$ for the first solution (`` + '' in Eq. (\ref{40})). Medium panel: the dependence of $V''(a_0)$ on the quantities $\omega$ and $Q^2/M^2$ (`` + '' in Eq. (\ref{40})). Right panel: the dependence of $a_0/M$ on the quantities $\omega$ and $Q^2/M^2$ for the second solution (`` - '' in Eq. (\ref{40})).}\label{f5}
\end{figure}
In summary, for the Schwarzschild-anti de Sitter case, the stable thin-shell wormholes constrained by cosmological observations must satisfy the following two conditions: (i) $\omega\in(-1,-0.89]$; (ii) $\Lambda M^2<F(a_0/M)$ when $2<a_0/M$ and $a_0/M\neq4.5$.

\subsection{Reinssner-Nordstr\"{o}m Thin-shell Wormholes}
For the Schwarzschild-anti de Sitter spacetime, the metric function $f(r)$ has the form $f(r)=1-2M/r-Q^2/r^2$, where $Q$ is the charge of the black hole. For $0<|Q|<M$ this geometry has two horizons situated at $r_{\pm}=M\pm\sqrt{M^2-Q^2}$. To form a thin-shell wormhole, we still need that the stable junction radius is larger than the outer horizon, i.e., $a_0>2M$. Then the potential $V(a)$ can be expressed as
\begin{equation}
V(a)=1-\frac{2M}{a}+\frac{Q^2}{a}-(1-\frac{2M}{a_0}+\frac{Q^2}{a_0})\left(\frac{a_0}{a}\right)^{4\omega+2}.  \label{38}
\end{equation}
Seeing $V(a_0)=0$, we also have from the condition $V'(a_0)=0$
\begin{equation}
\omega=-\frac{1}{2}\frac{(\frac{a_0}{M})^2-\frac{a_0}{M}}{(\frac{a_0}{M})^2-2\frac{a_0}{M}+\frac{Q^2}{M^2}}.  \label{39}
\end{equation}
Solving $a_0/M$, we obtain two real solutions as
\begin{equation}
\frac{a_0}{M}=\frac{1+4\omega\pm\sqrt{(1+4\omega)^2-8\omega(1+2\omega)(\frac{Q}{M})^4}}{2(1+2\omega)}.  \label{40}
\end{equation}
Utilizing the $1\sigma$ range $\omega\in[-1.05,0.89]$ and $0<|Q|/M<1$, we can investigate apparently the behaviors of $a_0/M$ and $V''(a_0)$. In the left panel of Fig. \ref{f5}, we show the dependence of $a_0/M$ on the quantities $\omega$ and $Q^2/M^2$, and find that $a_0/M$ is always satisfy the condition $a_0/M>2$. However, in the medium panel of Fig. \ref{f5}, we find that $V''(a_0)$ is always negative in the ranges $\omega\in[-1.05,0.89]$ and $0<|Q|/M<1$. Therefore, the first solution (use `` + '' in Eq. (\ref{40})) is unstable. Subsequently, in the right panel of Fig. \ref{f5}, we also show the dependence of $a_0/M$ on the quantities $\omega$ and $Q^2/M^2$ for the second solution (use `` - '' in Eq. (\ref{40})), and find that $a_0/M<2M$ in the ranges $\omega\in[-1.05,0.89]$ and $0<|Q|/M<1$. Hence, the second solution is also unstable.

In summary, for Reinssner-Nordstr\"{o}m case, the stable thin-shell wormholes constrained by cosmological observations do not exist.

\section{Discussions and Conclusions}
Since the elegant discovery that the universe is undergoing the phase of accelerated expansion, theorists have paid more and more attention to the renewed field---wormholes. Recently, with rapidly mounting cosmic data, we are full of interest in digging up the astrophysical information which hides in the cosmological observations.

In this study, we are dedicated to explore stable thin-shell wormholes constrained by modern cosmological observations. Without loss of generality, we study thin-shell wormholes constrained by observations in $\omega$CDM model. First of all, we place constraints on $\omega$CDM model using a combination of Union 2.1 SNe Ia data, the latest $H(z)$ data and CMB measurements. Subsequently, we use the constrained dark energy EoS $\omega$ which lies in $[-1.05,-0.89]$ to investigate thin-shell wormholes of different spacetimes, and obtain the following conclusions: the Schwarzschild thin-shell wormholes are unstable; the stable Schwarzschild-de Sitter thin-shell wormholes must satisfy two conditions, i.e., $a_0/M>15.45$ and $\omega\in[-1.05,-1)$; the stable Schwarzschild-anti de Sitter thin-shell wormholes also need two necessary conditions, i.e., $\omega\in(-1,-0.89]$ and $\Lambda M^2<F(a_0/M)$ when $2<a_0/M$ and $a_0/M\neq4.5$; the stable  Reinssner-Nordstr\"{o}m thin-shell wormholes constrained by cosmological observations do not exist.

One can easily find that the cosmological observations have provided a substantially tight constraint on the thin-shell wormholes in $\omega$CDM model. It is worth noting that, the method we developed in this study can be applied into exploring the stable thin-shell wormholes from any black hole spacetime in the framework of any cosmological theory. Furthermore, we have built a bridge between the cosmological observations and astrophysics: on the one hand, one can study the properties of celestial bodies by utilizing the cosmological observations, for instance, stable thin-shell wormholes in this work, and so forth; on the other hand, one can conversely extract the information at astrophysical scales to study the evolution of the universe at cosmic scales.

We expect that more and more high-quality data can bring us evolutionary cognition about various celestial bodies, for instance, black holes, wormholes, white dwarfs, pulsars, quasars, and so forth.

\section{acknowledgements}
The author Xin-He Meng warmly thank Professors Bharat Ratra and Sergei. D. Odintsov for helpful discussions on cosmology and astrophysics. The author Deng Wang thanks Prof. Jing-Ling Chen for beneficial feedbacks about quantum entanglement. The study is partly supported by the National Science Foundation of China.

\end{document}